# The chemistry induced by mechanical load

Irmgard Frank, Institut für Physikalische Chemie und Elektrochemie, Universität Hannover, Callinstr. 3A, Hannover

**Abstract**

Results in the field of mechanochemistry are summarized. For example under mechanical load often a solvolysis is observed instead of a simple homolytic bond rupture. Also a mechanically induced redox reaction was reported experimentally and could be modelled theoretically. With CPMD simulations it is possible to understand multi-step mechanisms in full detail.

**Introduction**

The term mechanochemistry started to become fashionable again about 15 years ago. Let me give a personal view of these past 15 years, summarizing and explaining from a distance some of our own papers. Mechanochemistry has an older story in the context of expressions like sonochemistry or tribochemistry. Originally the term mechanochemistry denoted the chemistry observed when solids are heavily grinded in a mortar. The renewed interest stems from the efforts of Hermann Gaub in Munich. He claimed he had measured the force necessary to break a covalent chemical bond, using a scanning tunneling microscope (STM). Results are in the region of 1 – 2 Nanonewton [1]. Just imagine to do that for a strand of a spider web, and then, way smaller, for a single covalent bond!  Breaking a single covalent bond within a polymer is certainly easy with any apparatus, but keeping it between the tip of a STM cantilever and a surface in a well-defined state, till the force is measured requires some art of experimentation.

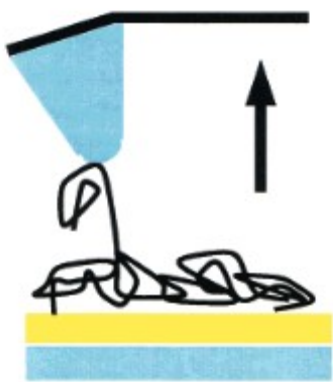

*Fig. 1: Experimental setup: A polymer fixed between the tip of a cantilever and a surface.*

**Results and Discussion**

Is it really possible to control a single molecule like this? Starting from 1998, we tried to understand these experiments using mainly Car-Parrinello molecular dynamics (CPMD) simulations. How would one theoretically approach the question how big the force must be to break a chemical bond? Chemists are used to energies, not to forces. How would one pictorialize what is happening? Let us look at the mechanically induced breaking of a single bond. There we have the Morse potential. By the mechanical load the system moves to larger distances on this Morse potential. The rupture force is found as the maximum derivative of the potential energy curve. Fine, this is simple. Any Hartree-Fock (HF) or density-functional theory (DFT) calculation will do for the accuracy needed. However, forces of 6 – 8 Nanonewton are obtained for single covalent bonds (about 80 - 100 kcal/mol binding energy). Hermann Gaub is convinced of his experiment, so why not try a little harder.

6 – 8 Nanonewton may be the right number for the breaking of a single isolated bond. The experimental setup is more complex. First, we do not expand a single bond, but many in a single polymer strand. Then the full system consists in a surface, a polymer which is bound to the surface, a cantilever attached to the polymer and a solvent droplet environing the polymer. We assume that the polymer was successfully strongly attached to both, the tip and the surface. The full polymer is expanded and will with certainty break when every bond within the polymer is pulled to a point where it gained 100 kcal/mol above the minimum of its Morse potential. If there are 10000 monomers in the chain, this makes 1000000 kcal/mol. Wait a moment! It will break way earlier and only a single or a few bonds will break. The polymer will not be pulverized as long as we pull with velocities well below the speed of sound. There are vibrations within the chain and a redistribution of energy. At a certain temperature we have a Maxwell-Boltzmann distribution of velocities within the chain. Some bonds will have more energy, some less. The chain might not break immediately when the total strand gained 100 kcal/mol since the energy is distributed, every single bond having just 0.01 kcal/mol on average. It will break some when in between. We derived a rather complicated formula for the rupture probability [2] which however can easily enough be treated with a pocket calculator. We used the assumption that the force depends quadratically on the deviation of the distance from the ideal conformation. This turned out to be a rather accurate assumption: The polymer breaks at a point where the harmonic region is not left yet. Using a Morse potential does not make the consideration way more accurate.

A harmonic potential means a linear force. It rises continuously till the rupture point is reached. At that point it falls suddenly to zero. The two rests of the molecule relax quickly to a constant value for the energy. That looks similar in static calculations and in molecular dynamics simulations. In static calculations the rupture point is dependent on the patience of the investigator to make all distances simultaneously bigger till the maximum rupture point is reached. In a molecular dynamics simulation the vibrational energy is distributed and the chain breaks way earlier. That all depends on the velocity with which the chain is expanded. This is a little unsatisfying. Is there something like an equilibrium velocity? Unfortunately no. The minimum velocity and minimum force at which the polymer chain breaks are zero. The polymer decomposes within a finite time, maybe years, also without stress.

There are essentially two possibilities to exert stress on a molecule in a molecular dynamics simulation: Constant velocity and constant distance. We mostly chose the first or also a combination: First a phase of expansion at constant velocity and then waiting at a constant distance. Not much to say about this choice, anything is allowed what is well-defined. We fail however to fully imitate the experimental conditions where a molecule is kept at constant force. This hardly ever influenced the interpretation, the constant distance mode being close enough to the constant force situation. We have way more problems with the limitations in size and simulation length. A constant velocity simulation was mostly the optimum solution, since it is easy to summarize: at this velocity and this temperature the polymer broke with that force. Particularly in solvent simulations where entropy plays a big role it was sometimes preferable to expand the chain, and keep it at this point till molecules from the environment interact, sometimes producing nice chemistry which I will come to below.

Around 2000 we also studied the question if one can, using mechanical load, reach the field of excited state chemistry, i.e. if one can bring organic dye molecules mechanically to the excited state or make it react as if it were in the excited state. Generally it can of course not be excluded. The systems we investigated, unsaturated carbon chains, however showed no excited state behavior. The hope had been that a double bond might isomerize under mechanical load. The finding is that the chains rather break in other positions than it would show this kind of excited state chemistry [2, 3]. In general, mechanochemistry is not at all similar to photochemistry.

We did not talk yet about the complexity of the whole system, I particular about the role of the solvent. The decision which bond breaks is way more difficult in a condensed phase environment. The weakest bond breaks, but which is the weakest bond if the system is exposed to the solvent? With Car-Parrinello molecular dynamics we can easily include a solvent due to the periodic boundary conditions used. We 'simply' fill up the simulation box with a solvent. Simply is inaccurate, it can mean quite some work for a patient PhD student in the case of more complex solvent molecules to get an equilibrated system with a realistic density. Equilibrated system means that the molecules obey a Maxwell-Boltzmann distribution which they do after some time of simulation. That sounds simple enough, but at the end of the equilibration one might not have the system anymore which one wanted to observe now under tensile stress. Input optimization can take some time for condensed systems. In experiment the solvent molecules have all the time in the world to attack the single stressed strand. From the point of view of a molecule which oscillates on a femtosecond time scale every potential reaction coordinate can be tested on the time scale of the experiment. In the simulation, however, we need to apply the stress way faster as we do not want to spend the electricity needed for a small town. We have to pull about six orders of magnitude faster than in experiment. Also at this velocity we succeeded to observe solvolysis of the chain and / or solvent reactions of the broken chain. The first case observed we published already in 2002 [4] and it is a quite complex case. Figures 2 to 5 tell the story. Polyethylene glycol under load normally breaks into radicals. This is not the case if the C-O bond can be polarized by surrounding solvent molecules. In that case, ions are formed in a heterolytic bond cleavage. These then quickly react with the solvent. In summary, a single water molecule is built into the chain, dissolving the chain in two pieces with OH groups. Talking about rupture forces on the basis of the CPMD calculations does not make so much sense in this case. The interaction with the solvent involves a big entropy contribution. The breaking and making of hydrogen bonds in the surrounding influences the potential energy curves too strongly. Far from the experimental pulling velocities a numerical comparison is not possible. This simulation is 15 years old. Today one would increase the size of the system. A much bigger system size is however not possible if one wants to pull slow enough, which means significantly slower than with the speed of sound which would cause other phenomena like complete fragmentation.

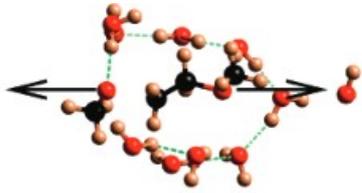

Fig. 2: A short piece of polyethylene glycol is expanded in the simulation

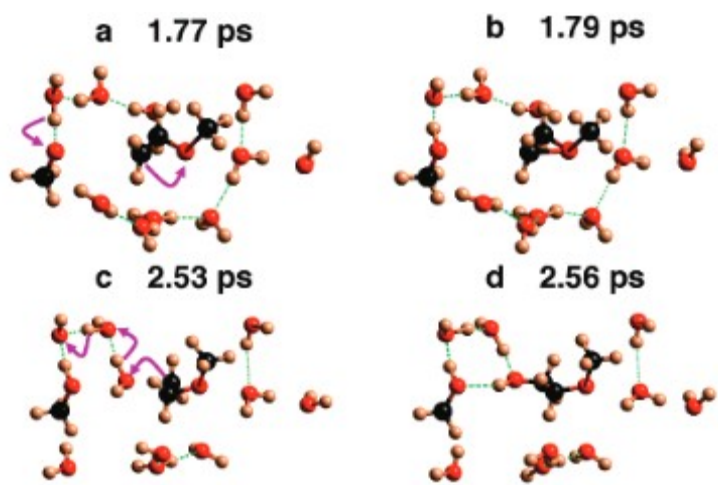

Fig. 3: Upon expansion, solvolysis occurs. In the end, after several proton transfers, in total a water molecule is built into the chain. The mechanism has acid-base character.

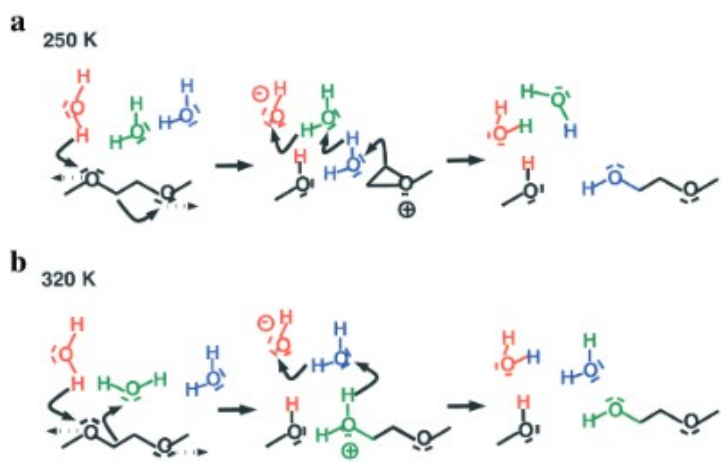

Fig. 4: Schematic description of two possible solvolysis pathways as observed in the simulations.

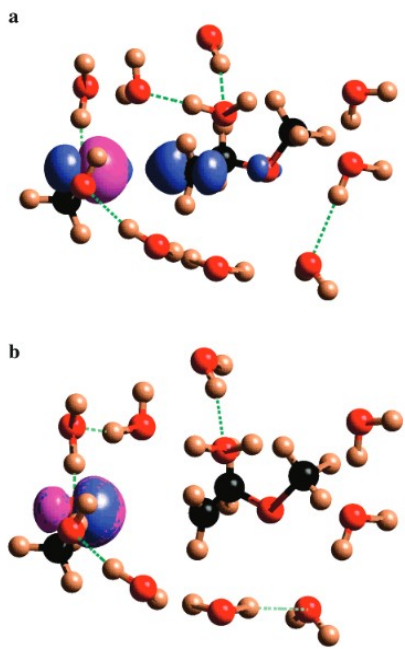

*Fig. 5: What the two orbitals (pink and blue) which form the C-O bond do during the reaction can be followed in full detail as simulated with CPMD. Influenced by the solvent the chain decides to break heterolytically with both electrons moving to the same side. In between some radical character is observed.*

Figure 5 shows what one can do with CPMD simulatios: We can follow in detail the motion of the electrons of the breaking bond. Initially there is a tendency to homolytic bond rupture: The two electrons are on different sides. Then a proton of a water molecule gets closer and succeeds to polarize the bond leading to heterolytic bond rupture and solvolysis.

Our effort in the field of mechanochemistry continued in a collaboration with Wacker Chemie AG [5, 6, 7, 8]. The attempt was to give an explanation for the effectivity of silicone additives in providing mechanical strength and hereby a direction for developing new additives. So we investigated the behavior of silicon dioxide under mechanical load. We found a ionic breaking of the bond which means that established models developed for homolytic C-C bond breaking do not apply [5]. In fact this finding explains the difference in mechanical stability between polyethylene and silicones, a difference which everyone knows. Compare a plastic bag and a piece of silicon rubber. The latter is easily indented with a finger nail and then broken into two pieces. This is not the case for the plastic bag. Here a C-C bond breaks homolytically, in the other case a Si-O bond breaks heterolytically. Both bonds are of similar strength and break at similar load. What makes the difference is the high reactivity of the ionic fragments. They react immediately with neighbouring bonds, way more aggressively than radicals [6, 8]. Till the ions are neutralized, one observes quick crack propagation. Additives must be able to neutralize these ions if that process shall be stopped. As part of this project we also studied an attachment to a surface and found not surprisingly that the attachment has a tendency to be weaker than the bonds around. In this study we observed rupture forces as small as 2 Nanonewton for a covalent bond breaking at an attachment [7]. Normally in the CPMD simulations we observed rupture forces of about 4 Nanonewton, probably due to too high pulling velocities. If in experiment mainly the rupture of the (covalent) attachment is observed, cannot be answered with certainty. There is also a high probability that one of the many bonds in the chain breaks. Also there is a high probability that knots are formed and like for a macroscopic rope, such points might be more prone to breaking.

In the past years we then investigated highly complicated reactions. Let us say, the reaction patterns seen before became too boring. Fernandez et al. described in a beautiful paper [9] a mechanically induced redox reaction in a disulfide bond breaking in the presence of a weak reducing agent (DTT, dithiothreitol). The simulation [10], shown in Figures 6 to 8, was probably less difficult than the experiment. Again we follow the motion of the electrons in full detail. We can add a nice picture describing the similarity and difference between this redox reaction and previously observed solvolysis reactions which rather have the character of acid-base reactions (Fig. 9).

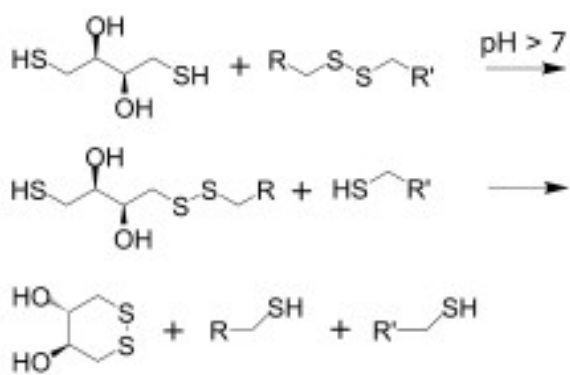

Fig. 6: Dithiothreitol is a weak reducing agent for disulfide bonds. The second reaction step stabilizes the product state due to entropy. Also mechanical load can shift the equilibrium.

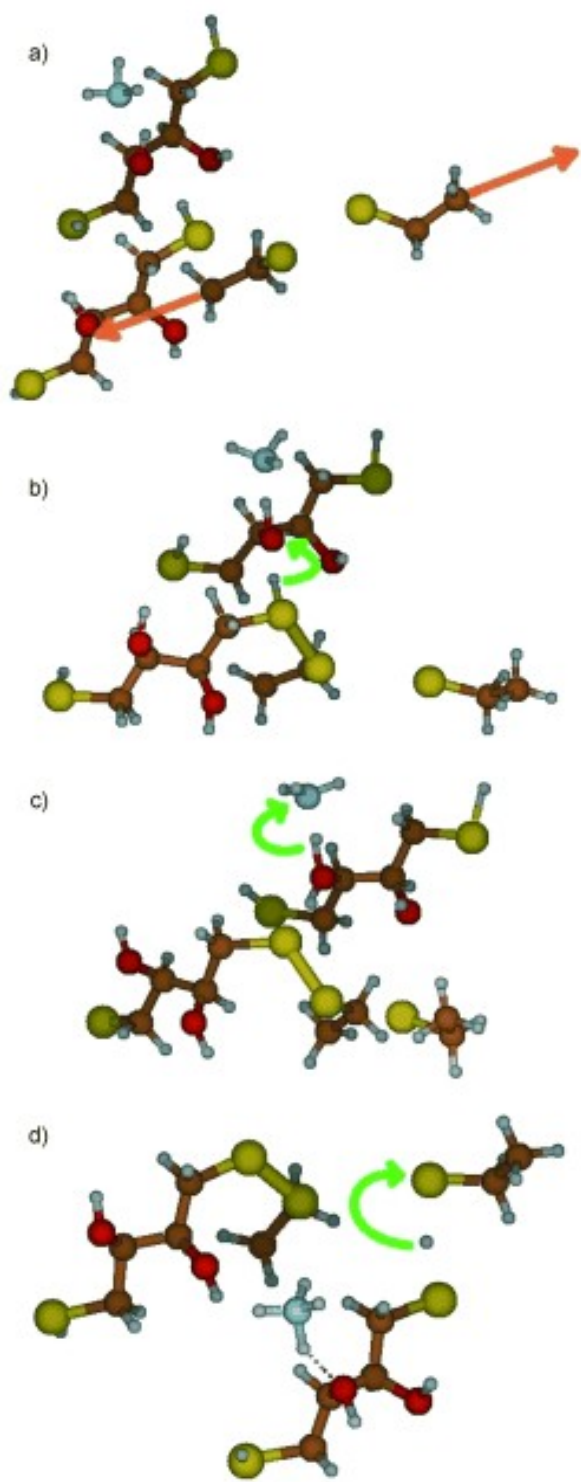

*Fig. 7: Dithiothreitol attacks under mechanical load. Already in the first picture the breaking S-S bond is strongly expanded. The following pictures show proton transfers which occur after the DTT attack.*

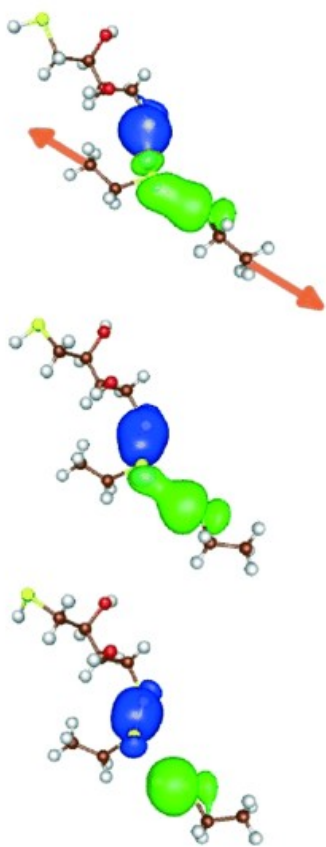

*Fig. 8: The motion of the relevant orbital pairs (altogether four electrons, two shown in blue, two in green) during the mechanically induced reaction. They electrons stay paired during the whole reaction, which is rather unusual for a redox reaction.*

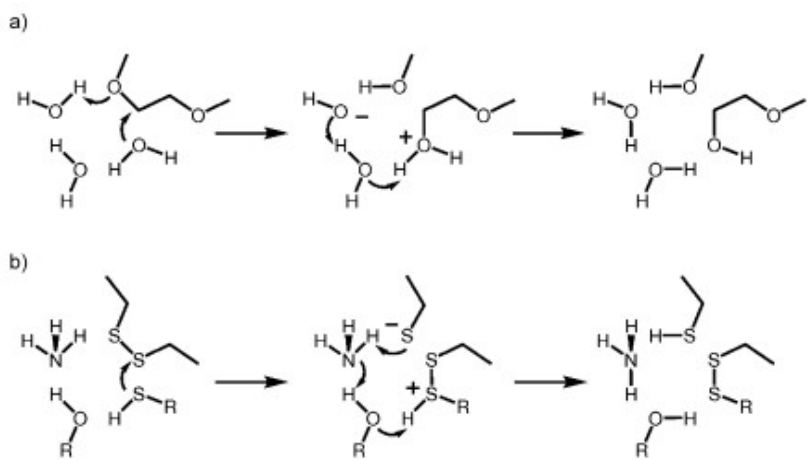

*Fig. 9: Scheme illustrating differences and similarities between the before observed solvolysis reactions and the redox reaction. The mechanical stress leads to the formation of ions as short-lived intermediates which immediately disappear again in a stabilizing reaction. In both cases one molecule is built into the chain.*

**Outlook**

What is the future of mechanochemistry? Apart from standardization of the existing results one could think of applying pressure in a molecular dynamics simulation. We made some attempts and it is certainly clear that extreme pressure is needed, otherwise the systems succeed to avoid breaking a covalent bond. In an experiment one will usually not deal with single crystals, then normally the breaking at grain boundaries contributes strongly. Nevertheless this is certainly a field to investigate in the future years. Think of the basis of friction. If two perfect metal surfaces are perfectly put at each other, then they form one single metal block, a phenomenon which is called adhesion. Normal surfaces are not so perfect, experience reconstruction, and are oxidized and hydroxylated. Then water and other solvents play a big role in friction and adhesion. Under extreme mechanical load, water might hydrolyse a rugged surface, a lubricant might decompose. Lots of phenomena still to be investigated in the field of mechanochemistry!

Table of Contents

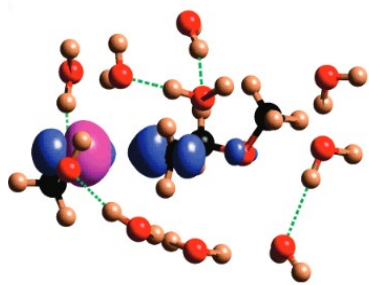

Heterolytic or homolytic? If mechanically expanded in a solvent environment slow enough, acid-base reactions or redox reactions occur in contrast to radical reactions. Car-Parrinello molecular dynamics simulations show the full complexity. A simple scheme for the mechanism of such reactions is derived.